\documentstyle[11pt]{article}
\begin{document}
\begin{center}
{\Large \bf Properties of $^{8}$Be and $^{12}$C
 deduced from the folding--potential model}\\
[0.7cm]
P.~Mohr, H.~Abele, V.~K\"olle, G.~Staudt\\
{\small\it Physikalisches Institut, Univ.~T\"ubingen, D--72076 T\"ubingen,
Germany}\\
[0.3cm]
H.~Oberhummer and H.~Krauss\\
{\small\it Institut f\"ur Kernphysik, Technische Universit\"at Wien,
A-1040 Vienna, Austria}\\
\end{center}

{\bf Abstract:}
  The $\alpha$--$\alpha$ differential cross sections are analyzed in the
optical
  model using a double--folded potential. With the knowledge of this potential
bound and
  resonance--state properties of $\alpha$--cluster states in $^{8}$Be and
$^{12}$C as well
  as astrophysical S--factors of $^{4}$He($\alpha$,$\gamma$)$^{8}$Be and
  $^{8}$Be($\alpha$,$\gamma$)$^{12}$C are calculated. $\Gamma_{\gamma}$--widths
and
  B(E2)--values are deduced.


In order to determine reaction rates for the triple--alpha process \cite{Obe0},
we have calculated in the potential model the $\alpha$+$\alpha$
differential cross section as well as the bound and resonance--state
properties of $\alpha$--cluster states in  $^{8}$Be and $^{12}$C.
The potentials are calculated using the folding procedure \cite{Kob},
\cite{Obe1}.

In this approach the nuclear densities
are derived from nuclear charge distributions \cite{Vri} and folded
with an energy and density dependent NN--interaction
$v_{\rm eff}$ \cite{Kob}. By means of a normalization factor $\lambda$
the depth of the potential can be adjusted to elastic scattering
data and to bound and resonant
state energies of nuclear cluster states.
In this work the folding potentials were determined using the
computer code DFOLD \cite{Abe1}. The imaginary part of the optical
potential can be neglected at the low energies considered.

Using the folding procedure for the $\alpha$+$\alpha$
scattering potential we fitted the experimental
differential cross section for all projectile energies
below the break--up threshold ($E_{\rm lab} \approx 7$--$35$\,MeV)
with a renormalization factor $\lambda = 1.656$.
An excellent agreement between the calculated
and experimental cross section has been found \cite{Abe2}.
With the same potential we reproduce the resonance energies and widths of
the $^{8}$Be ground--state: $E_{\rm R}^{\rm th} = 92.1$\,keV and of
the $2^{+}$-- and $4^{+}$--state in $^{8}$Be (see Table 1).
Especially for the ground state the calculated values agree excellently
with the experimental data. This result confirms that this nucleus can
be well described in a two-cluster approach \cite{Wil}.
This marked $\alpha$--cluster structure is also
reflected by the values of the $\alpha$--particle
spectroscopic--amplitudes for these $^{8}$Be--states \cite{Kur}:
{\cal S}(g.s.)=1.48, {\cal S}($2^+$)=1.46, and
{\cal S}($4^+$)=1.39.

The s--wave function for the relative motion of the two $\alpha$--clusters in
the $^8$Be
ground state can be folded with the nucleon densities of both
$\alpha$--particles to
calculate the $^8$Be density \cite{Abe2}. In our calculation we used an upper
cut-off radius of 20 fm.
Averaging over 4$\pi$ we obtain the nucleon density distribution. The result is
shown in
Fig.\ 1. With the knowledge of the $^8$Be density distribution we are able to
calculate the
optical potential for the $\alpha$--$^8$Be system using the folding procedure
again. Since
no experimental phase shifts for $\alpha$--$^8$Be are available, we have to fit
the
normalization factors $\lambda$ of the folding potentials to the bound and
resonance--state
energies of $^{12}$C.

The ground state ($0_1^+$), the first ($2_1^+$) and second ($0_2^+$) excited,
and
the ($2_2^+$) state in   are those states which are of interest in the
triple--alpha process (see Table 2). With respect to a $\alpha \otimes ^8$Be
clustering
in $^{12}$C, the $0_1^+$ and $2_1^+$ bound states belong to the ground state
band with a
harmonic oscillator quantum number Q=4, whereas for the $0_2^+$-- and
$2_2^+$--state we assume
that they can be interpreted as members of a band with Q=6. We assume the broad
10.3 MeV
level in $^{12}$C to be the $2_2^+$ state. The adopted spin assignment of this
level is
$0^+$ \cite{Ajz2}. However, an assignment of $2^+$ for this state is strongly
suggested by
experimental results \cite{Jac1} as well as by microscopic calculations
\cite{Fuj}.

\begin{table}
{\bf Table 1.} Relevant data of resonant states in $^{8}$Be
$$\vbox{
\halign {\hfil # \hfil \quad & \hfil # \hfil\quad & \hfil # \hfil\quad & \hfil
# \hfil\quad  &\hfil # \hfil \quad\cr
\noalign{\hrule\smallskip}\cr
\ $J^{\Pi}$\hfil & $E^{\rm calc}$ [MeV]\hfil & $E^{\rm exp}$ [MeV]
\cite{Ajz1}\hfil &
$\Gamma_{\alpha}^{\rm calc} $ [eV]\hfil & $\Gamma_{\alpha}^{\rm exp} $[eV]
\cite{Ajz1}\hfil \cr
\noalign{\smallskip\hrule\smallskip}\cr
\ $0^+$ & 0.0921 & 0.09212(5) & 6.86 & $6.8\pm 1.7$ \cr
\ $2^+$ & 2.78 & 3.13(3) & $1.3\cdot 10^6$ & $1.5(2) \cdot 10^6$ \cr
\ $4^+$ & 11.02 & 11.5(3) & $ ^{>}_{\sim}3\cdot 10^6$ & $\approx 3.5\cdot 10^6$
\cr
\noalign{\smallskip\hrule} \cr } }$$
\end{table}
\begin{table}
{\bf Table 2.} Relevant data of bound and resonant states in $^{12}$C
$$\vbox{
\halign {\hfil # \hfil \quad & \hfil # \hfil\quad & \hfil # \hfil\quad & \hfil
# \hfil\quad  &\hfil # \hfil \quad\cr
\noalign{\hrule\smallskip}\cr
\ $J^{\Pi}$\hfil & $E$ [MeV]\hfil & $\lambda$ \hfil &
$\Gamma_{\alpha}^{\rm calc} $ [eV]\hfil & $\Gamma_{\alpha}^{\rm exp} $[eV]
\cite{Ajz2}\hfil \cr
\noalign{\smallskip\hrule\smallskip}\cr
\ $0_1^+$ & 0.0 & 1.144 & --- & --- \cr
\ $2_1^+$ & 4.437 & 1.010 & --- & --- \cr
\ $0_2^+$ & 7.654 & 1.286 & 7.5 & 8.3$\pm$1.0 \cr
\ $2_2^+$ & 10.3 & 1.152 & $\approx 2.0\cdot 10^6$ & (3.0$\pm$0.7)$\cdot 10^6$
\cr
\noalign{\smallskip\hrule} \cr } }$$
\end{table}
\begin{table}
{\bf Table 3.} $\Gamma_{\gamma}$--widths and B(E2)--values
$$\vbox{
\halign {\hfil # \hfil \enspace & \hfil # \hfil\enspace & \hfil # \hfil\enspace
& \hfil # \hfil\enspace  &\hfil # \hfil
\enspace
&\hfil # \hfil \enspace&\hfil # \hfil \enspace\cr
\noalign{\hrule\smallskip}\cr
\ \hfil & $E_{\rm f}$ [MeV]\hfil & $J^{\pi}_f$ \hfil &
$E_i$ [MeV]\hfil & $J^{\pi}_{\rm i}$\hfil & $\Gamma _{\gamma}$ [meV]\hfil
& B(E2) [W.u.] \hfil \cr
\noalign{\smallskip\hrule\smallskip}\cr
\ $^8$Be & 0.0 & $0^+$ & 3.04 & $2^+$ & 8.4 & 75 \cr
\ $^{12}$C & 4.437 & $2_1^+$ & 7.654 & $0_2^+$ & 4.1 & 8.8 \cr
\  & 7.654 & $0_2^+$ & 10.3 & $2_2^+$ & 17 & 100 \cr
\noalign{\smallskip\hrule} \cr } }$$
\end{table}

Using the double--folded $\alpha$--$^8$Be potential we have fitted the energies
of these
$^{12}$C states. These states have quite a different $\alpha \otimes
^8$Be(g.s.)
cluster structure as can be seen from the values of the spectroscopic factors:
{\cal S}(g.s.)=0.557 \cite{Kur}, {\cal S}($2_1^+$)=0.154 \cite{Kwa}, and
{\cal S}($^{12}$C, $0_2^+$)=1.8 \cite{Fuk}. Therefore, the $\alpha$--$^8$Be
folding potentials have to be renormalized separately to the
ground and excited states in $^{12}$C.
The normalization factors $\lambda$ for the potential obtained in this
way are listed in Table 2. Using the optical model we also calculated the alpha
widths
$\Gamma _{\alpha}$ of the resonant $0_2^+$ and $2_2^+$ states. The good
agreement between
the experimental and the calculated $\alpha$--width for the
$0_2^+$--state confirms the assumption that this state has
a marked $\alpha \otimes ^8$Be(g.s.)--structure.

Furthermore, we calculated the cross sections and astrophysical $S$-factors of
the reactions
$^4$He($\alpha$,$\gamma _0$)$^8$Be and $^8$Be($\alpha$,$\gamma _1$)$^{12}$C in
the direct
capture model \cite{Obe1}, \cite{Moh}. These cross sections are essentially
determined by the overlap of the
scattering wave functions in the entrance channel, the bound state wave
functions in the
exit channel, and the electric quadrupole operator. The wave functions were
calculated using
the $\alpha$--$\alpha$ and $\alpha$--$^8$Be potential, respectively. In order
to compute the
reaction $^4$He($\alpha$,$\gamma _0$)$^8$Be the resonance energy of the $^8$Be
ground state
was lowered by about 100 keV. This means that this state was assumed to be
weakly bound.
For the description of the bound states the $\alpha$--particle spectroscopic
factors
given above are used. In Fig.\ 2 the
cross section of the resonant E2--capture into the $^8$Be ground state is
shown.
Excellent agreement is found with the result of a calculation in
a bremsstrahlung model \cite{Lang}.
Approximating the $\sigma (E)$ cross sections near the resonance energy by a
Breit-Wigner
parametrization, $\Gamma _{\gamma}$--widths and B(E2)--values can be deduced.
These
quantities are given in Table 3.

Acknowledgments: We want to thank the DFG--project Sta290/2, and the
Austrian Science Foundatio (FWF) (project P8806-PHY).

\newpage
\section*{Captions to figures}
\indent

Fig.~1: Nucleon density distribution of $^8$Be calculated in the potential
model

Fig.~2: Capture cross section of $^4$He($\alpha$,$\gamma _0$)$^8$Be
        near the 2$^+$ resonance

\end{document}